\documentclass{nseJournal}

\begin{document}

\title{Tritium accumulation and ozone decontamination of tungsten and beryllium}

\addAuthor{\correspondingAuthor{Dominic Batzler}}{a}
\correspondingEmail{dominic.batzler@kit.edu}

\addAuthor{Robin Größle}{a}
\addAuthor{Philipp Haag}{a}
\addAuthor{Elizabeth Paine}{b}
\addAuthor{Marco Röllig}{a}
\addAuthor{Marie-Christine Schäfer}{a}
\addAuthor{Marius Schaufelberger}{a}
\addAuthor{Kerstin Trost}{a}

\addAffiliation{a}{Tritium Laboratory Karlsruhe, Institute for Astroparticle Physics,\\ Karlsruhe Institute of Technology, Hermann-von-Helmholtz-Platz 1,\\ 76344 Eggenstein-Leopoldshafen, Germany}
\addAffiliation{b}{Department of Applied Physics, Eindhoven University of Technology,\\ 5612 AZ Eindhoven, The Netherlands}

\addKeyword{Tritium}
\addKeyword{Tritium Laboratory Karlsruhe}
\addKeyword{Tungsten}
\addKeyword{Beryllium}
\addKeyword{Beta-induced X-ray Spectrometry}

\titlePage

\begin{abstract}
Tritium adsorption on surfaces creates a variety of issues, ranging from the fields of fusion applications to small and large-scale laboratory experiments using tritium. The extent to which tritium accumulates on surfaces is generally material-dependent and must be determined through experiments. Additionally, this surface contamination necessitates the implementation of appropriate decontamination procedures, preferably in-situ. A suitable method could be exposure to ozone during UV irradiation. However, it is currently not known if both components are necessary for the decontamination. At Tritium Laboratory Karlsruhe, both questions on contamination and decontamination can be addressed using a single experimental setup. With this, it is possible to expose solid samples to gaseous tritium to measure the temporal activity evolution. Furthermore, the system can be filled with dry air, and dry air containing ozone to explore their decontamination effect. Both measurement modes were applied to beryllium and tungsten samples, which were chosen for their relevance in fusion. The beryllium surface was observed to accumulate tritium more than four times faster than tungsten when exposed to gaseous tritium. Concerning the decontamination, without simultaneous UV irradiation, exposure to ozone did not have any distinct effect on the surface activity compared to simply using dry air. This leads to the conclusion that UV illumination of the surfaces is required to achieve a significant decontamination factor.
\end{abstract}

\section{Introduction}
\label{sec:introduction}
Exposure to gaseous tritium often results in activity accumulation on the surface. Such contamination poses multiple challenges for a fusion fuel cycle, and tritium handling facilities in general. These include the loss of process gas, the accumulation of tritium inventory, the complication of tritium accountancy calculations, and the need for increased safety precautions during maintenance. In tritium monitoring systems, this undesired contamination results in a history-dependent, elevated background signal, which is often referred to as memory effect \cite{nishikawa89memory}. For neutrino mass experiments, such as the Karlsruhe Tritium Neutrino (KATRIN) experiment \cite{katrin21tdr,katrin25science}, this creates a systematic effect that needs to be studied and accounted for.

Therefore, suitable decontamination strategies, ideally in-situ, are required to mitigate the aforementioned issues. While traditional methods, such as bake-out, are somewhat successful at reducing the surface activity \cite{batzler25tungsten}, the high temperatures in this case could create additional problems such as increased tritium permeation.

UV/ozone exposure presents itself as a promising candidate for in-situ decontamination, which was already proven effective on tritiated surfaces, with a decontamination factor of up to three order of magnitudes \cite{krasznai95uv,aker24uv}. However, from literature it is not clear which component is responsible for the decontamination efficiency. Some reports suggest that ozone or oxygen radicals alone are sufficient \cite{ishikawa97ozone,li22oxygen}, others conclude that only the combination of ozone and simultaneous UV irradiation are effective \cite{aker24uv,vig85uv}. In case only ozone were required, the in-situ decontamination would be less complicated to put into practice.

To address these questions, the Tritium Activity Chamber Experiment (TRACE) was employed \cite{batzler25tungsten}. It is suited to expose solid samples both to gaseous tritium for contamination, and ozone (without UV irradiation) for decontamination. Both measurement modes were applied to tungsten and beryllium due to their significance for fusion. Beryllium is also of interest for use in the KATRIN experiment. In this work, the results on the surface activity evolution of tungsten and beryllium due to exposure to tritium, as well as the effect of ozone on their surface activity are reported.

\section{Experimental Setup}
\label{sec:exp_setup}

The TRACE system consists of a sample cell based on DN 40 CF components with optimised geometry to minimise tritium inventory. A gold-coated beryllium window serves as one of the interfaces and enables in-situ measurements of the sample's surface activity by beta-induced X-ray spectrometry (BIXS) \cite{matsuyama98bixs} . For A more detailed description of the experimental setup, see \cite{batzler25tungsten}. A UV-based ozone generator, similar to that used in \cite{batzler24uv} was connected to the measurement cell to expose the samples to ozone. Throughout the measurement campaigns, the purity of the tritium used for contamination was above 95\%.

\subsection{Sample preparation}

\begin{figure}
  \centering
  \begin{subfigure}[b]{0.3\textwidth}
    \centering
    \includegraphics[width=\textwidth]{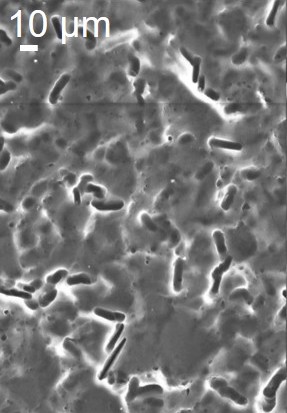}
    \caption{}
    \label{fig:beryllium_sem}
  \end{subfigure}
  \begin{subfigure}[b]{0.3\textwidth}
    \centering
    \includegraphics[width=\textwidth]{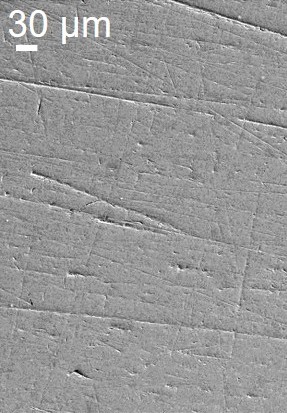}
    \caption{}
    \label{fig:tungsten_sem}
  \end{subfigure}
  \caption{SEM images of the beryllium (a) and tungsten (b) samples.}
  \label{fig:samples_sem}
\end{figure}

In this study, three S-65 beryllium samples by Materion with a minimum beryllium content of 99.2\% were used. These were left unpolished and were cleaned with ethanol prior to being installed in the measurement cell. A tungsten sample was cut from a rolled metal sheet with an elemental purity of over 99\% according to X-ray fluorescence analyses. Unlike the beryllium samples, the tungsten sample was manually polished and cleaned in an ultrasonic bath. \autoref{fig:samples_sem} shows images of the beryllium and tungsten surfaces taken using a scanning electron microscope (SEM), which reveal their different topologies.

Before the start of the measurement campaigns, the samples were baked out at 200°C until a combined leak and out-gassing rate of lower than $10^{-9}$\,mbar$\cdot$l/s was achieved.

\section{Results}
\label{sec:results}

\subsection{Contamination}

\begin{figure}[t]
  \centering
  \includegraphics[width=\linewidth]{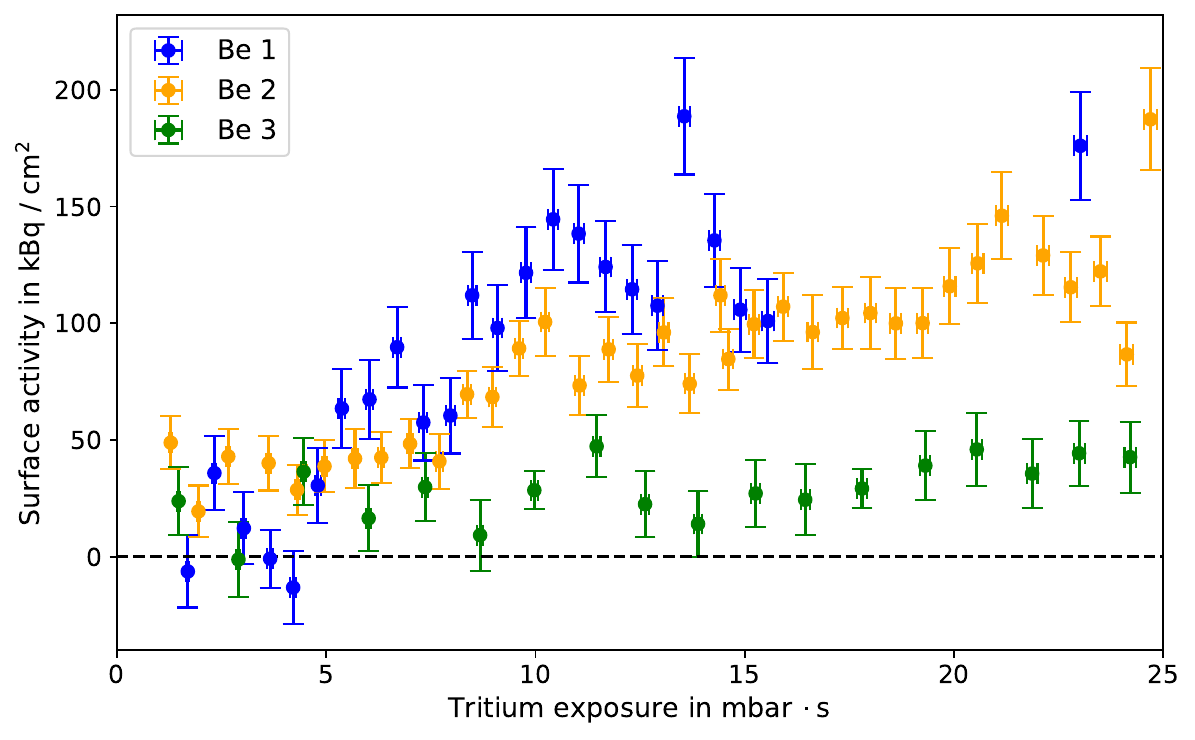}
  \caption{Increase of surface activity of three beryllium samples due to repeated exposure to gaseous tritium.}
  \label{fig:be_isotherms}
\end{figure}

The samples were exposed to gaseous tritium for 10\,min, followed by an evacuation period to measure the remaining tritium on the samples' surfaces in-situ using BIXS. By repeating this process, the temporal evolution of the samples' surface activity can be measured.
The exposure pressures were selected such that the tritium was in the free molecular flow regime in order to avoid effects from gas self-interaction.
For two beryllium and a tungsten sample, a pressure of $10^{-3}$\,mbar was chosen, while a third beryllium sample was exposed to $2\cdot 10^{-3}$\,mbar of tritium. Given the system's geometry, these pressures correspond to Knudsen numbers of 30 and 15, respectively.

The time series of the beryllium surface activities are shown in \autoref{fig:be_isotherms}. Both statistical and systematic uncertainties are included in the individual data points; the largest contribution stems from the uncertainty on the sample cell volume for the activity calibration. With each subsequent sample, less activity is accumulated, ranging from a slope of $(10.06\pm 0.70)$\,kBq/(cm$^2$\,mbar$\cdot$s) to $(2.28\pm 0.22)$\,kBq/(cm$^2$\,mbar$\cdot$s) for the first and third samples, respectively.

\begin{figure}[t]
  \centering
  \includegraphics[width=\linewidth]{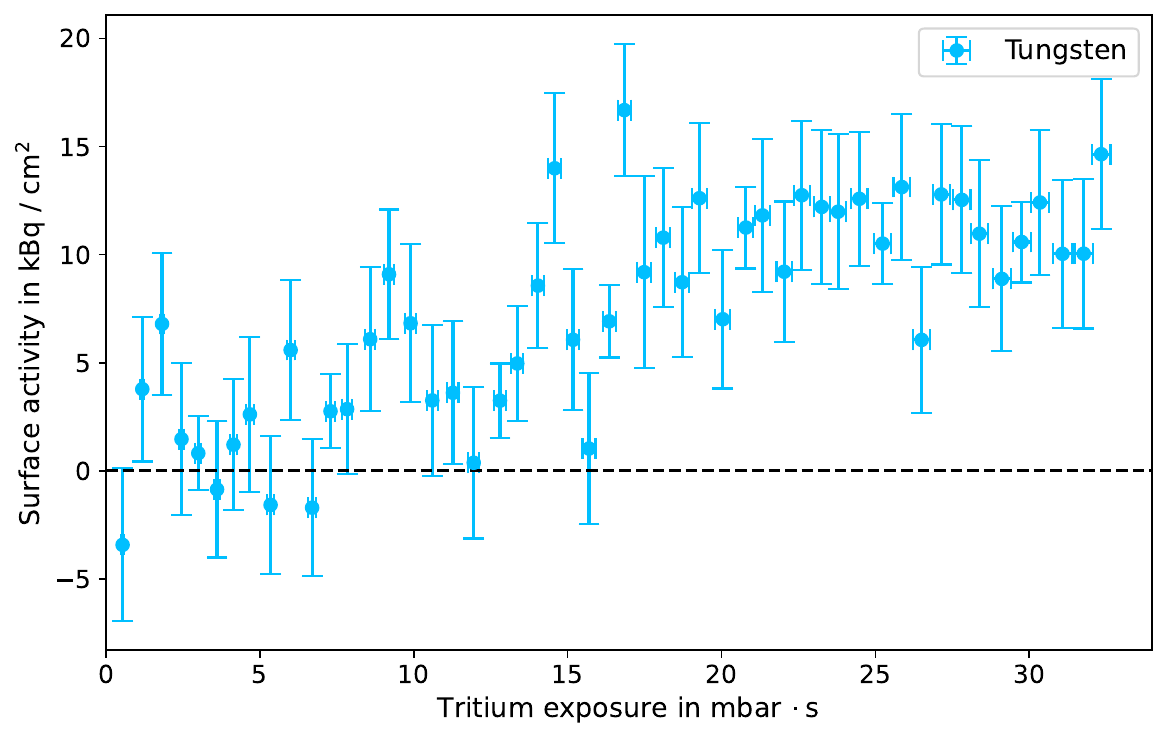}
  \caption{Surface activity evolution of the tungsten sample.}
  \label{fig:w_isotherm}
\end{figure}

With a slope of $(0.488\pm 0.026)$\,kBq/(cm$^2$\,mbar$\cdot$s), the tungsten sample retains considerably less tritium on its surface than beryllium, as displayed in \autoref{fig:w_isotherm}. Due to its higher nuclear charge of $Z=74$ compared to beryllium's $Z=4$, the uncertainties are also significantly lower.

\subsection{Ozone decontamination}

After concluding the contamination measurements, the third beryllium and the tungsten samples were used to test the ozone decontamination method. To investigate the impact of ozone, the sample was exposed to dry air taken from the glove box with and without prior UV irradiation. If the air was irradiated, the ozone content should be around 200\,ppm after expansion from the ozone generator to the sample chamber \cite{batzler24uv}. In both cases, the air/ozone were kept in the recipient for 20\,min to allow the ozone to react with the surfaces, and to have a consistent procedure \cite{batzler24uv}. Similar to the contamination measurements, multiple cycles of exposure and evacuation were applied.

\begin{figure}[t]
  \centering
  \includegraphics[width=\linewidth]{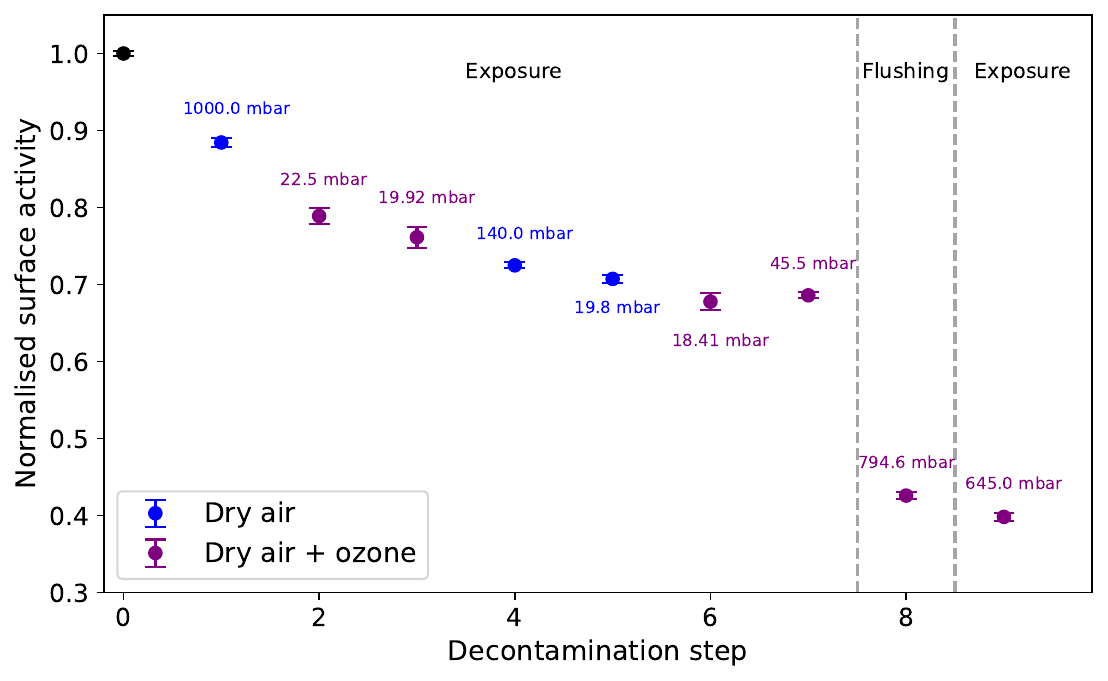}
  \caption{In-situ decontamination of the third beryllium sample with the corresponding pressure during exposure.}
  \label{fig:be_deco}
\end{figure}

\autoref{fig:be_deco} shows the influence of the decontamination procedure on the beryllium sample's surface activity, normalised to its starting value. The exposure pressures used for the individual data points are noted in the graph. Breaking the vacuum to connect the ozone generator for the first time was included as the first data point, since this already caused the surface activity to decrease by roughly 10\%. Two decontamination steps at low pressure using ozone-containing air, two steps with only dry air, and two steps with ozone followed. Each of these was comparable in decontamination efficiency, resulting in an activity of 70\% of the initial value after the six steps. The first significant reduction in activity could be achieved after introducing ozone-containing air at a rate of roughly 0.4\,mbar/s, until a pressure of roughly 800\,mbar was reached. A subsequent, rapid exposure at a relatively high pressure only had a marginal effect on the surface activity. Due to experimental boundary conditions, the decontamination measurements were concluded, with an overall achieved activity reduction of 60\%.

\begin{figure}[t]
  \centering
  \includegraphics[width=\linewidth]{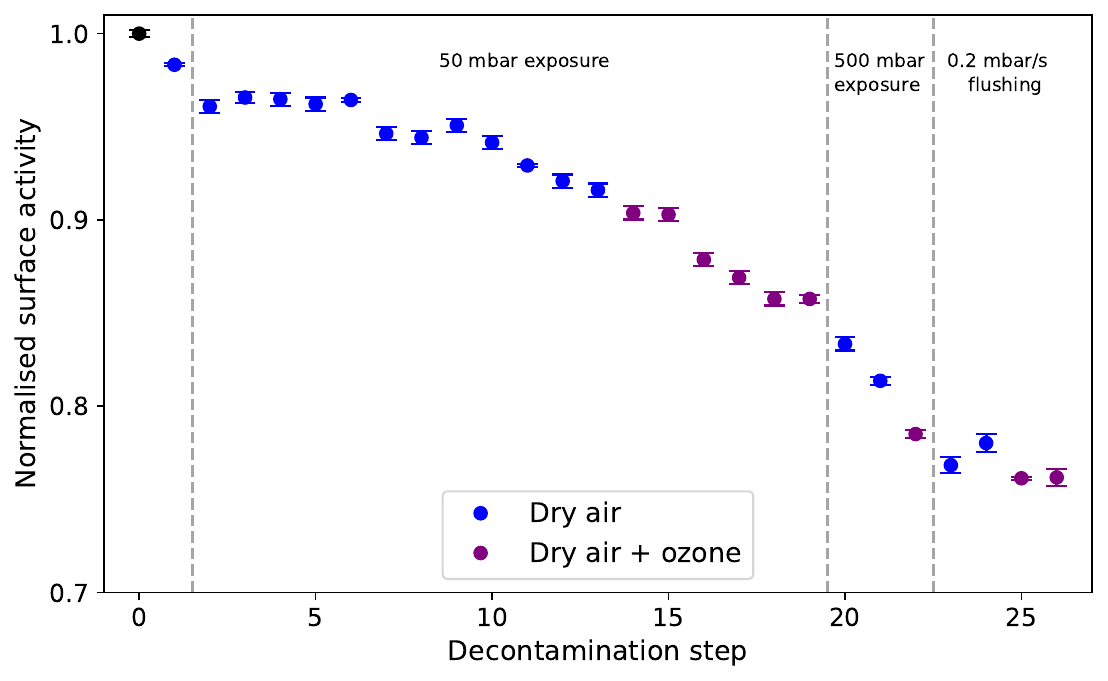}
  \caption{In-situ decontamination of the tungsten sample.}
  \label{fig:w_deco}
\end{figure}

The same procedure was applied to the tungsten sample, as seen in \autoref{fig:w_deco}. After installing the ozone generator, multiple exposure at 50\,mbar were carried out; first with dry air without, then with ozone. Each individual step removed up to 3\% of the initial surface activity, and no distinction between the absence or presence of ozone could be made. This also applies when higher pressures, in this case, 500\,mbar, are used. For the last four measurements, the pressure was slowly increased by 0.2\,mbar/s. These did not have any significant decontamination effect, which is different for the beryllium sample. All in all, about a quarter of the initial activity could be removed from the surface of the tungsten sample before finishing the measurements.

\section{Discussion}
\label{sec:discussion}
Each beryllium sample accumulated less tritium on its surface than the previous one. A reason for this could be the fact that most of the components in the vicinity of the sample were in pristine condition before the start of the measurements. As the samples were exposed to tritium, the surrounding surfaces could have been gradually saturated, such that their tritium adsorption decreased over time, until reaching a stable level. Nevertheless, the tungsten sample retained significantly less tritium on its surface, namely less than a quarter compared to beryllium. It is unlikely that this observation can be explained by the differences in surface structure, or rather number of adsorption sites, since their final activities correspond to surface coverages below 1\%. 

For the decontamination measurements, using dry air with and without ozone reduces the samples' surface activity by comparable amounts. The most probable cause for the tritium desorption is an initial flushing effect at the start of the exposure, for which the ozone is irrelevant.

\section{Conclusion and Outlook}
\label{sec:conclusion}
With the TRACE setup, it is possible to investigate tritium accumulation, as well as decontamination methods using different gases in-situ. This was demonstrated using three beryllium samples and a tungsten sample. Assuming exposure to gaseous tritium in the molecular flow regime is the primary cause of adsorption, beryllium accumulates tritium at a faster rate than tungsten.
From the decontamination measurements presented in this work, it is not possible to confirm a beneficial effect of ozone in dry air without UV illumination. If this turns out to be true, it will potentially be more complicated to carry out an in-situ decontamination on larger tritiated systems, since not only ozone is required, but the surfaces need to be irradiated by UV light simultaneously.

Designing the next iteration of the system is currently in progress. The upgraded setup will have the capability to heat up the sample to up to 1500°C, e.g. to reproduce the conditions in a fusion fuel cycle more accurately. Additionally, it will be possible to directly irradiate the sample with UV light. This way, it will also be possible to investigate the combined in-situ UV/ozone decontamination, which is required to realise and establish the method on a more technical scale.

\pagebreak
\section*{Acknowledgements}
The authors would like to thank the TRIHYDE and CAPER teams at Tritium Laboratory Karlsruhe, who provided the tritium to carry out this study, and who purified the resulting process gas, respectively.

\pagebreak
\bibliographystyle{ans_js}                                                                           
\bibliography{ref.bib}

\end{document}